\documentclass[10pt]{article}
\usepackage[utf8]{inputenc}
\usepackage[T1]{fontenc}
\usepackage{amsmath}
\usepackage{amsfonts}
\usepackage{amssymb}
\usepackage[version=4]{mhchem}
\usepackage{stmaryrd}
\usepackage{bbold}
\usepackage[english]{babel}
\usepackage[a4paper, left=3cm, right=3cm, top=2cm]{geometry}
\usepackage{graphicx}
\usepackage{xltabular}
\usepackage[]{hyperref} 
\hypersetup{
    colorlinks=true,
    linkcolor=blue,
    citecolor=blue,
    filecolor=magenta,      
    urlcolor=blue,
    pdftitle={Overleaf Example},
    }
\date{}
\graphicspath{ {./images/} }

\newtheorem{theorem}{Theorem}[section]
\newtheorem{definition}[theorem]{Definition}
\newtheorem{lemma}[theorem]{Lemma}

\newtheorem{remark}[theorem]{Remark}
\newtheorem{corollary}[theorem]{Corollary}

\DeclareUnicodeCharacter{0131}{$\imath$}

\begin{document}
\title{Construction of envelopes of holomorphy and QFT}

\author{Ulrich Armbrüster}
\maketitle

\textbf{Abstract. }  Methods of continuation of holomorphic functions of several complex variables are investigated within the axiomatic framework of Araki, Haag, and Kastler in local quantum field theory. The motivation comes from the analysis of a mass gap in an energy-momentum spectrum without vacuum vector. The main conclusion is some non-restrictedness property in a mass gap situation. Prior to that, some results on holomorphic functions related to a mass-gap-situation are obtained and investigated.

\tableofcontents

\section{Introduction}\label{Introduction}
In this article we are discussing investigations on the energy-momentum spectrum within the algebraic formulation of quantum field theory. Unlike the Wightman approach, only observables represented by bounded operators play a role in the algebraic formulation. This limitation is not restrictive since instead of the usual self-adjoint operators only their spectral projectors, which are always in the closure of the representation of the observable algebra, are considered. Even in the slightly more general case of a closeable operator, one can still consider only bounded operators by polar decomposition of the closure into an isometric and a self-adjoint operator. This approach makes it mathematically easier to handle the operators as one does not need to pay attention to their respective domains.\

The fundamental physical principles of the algebraic approach are isotonicity, translation invariance, Einstein causality, and the spectral condition. The axioms required in this article are described in section \ref{QFT} and are based on the works of Araki, Haag, and Kastler in \cite{HaKa} and \cite{Ara}.\

Starting with these axioms together with the spectrum condition for the energy-momentum operator, one can construct functions with holomorphic Fourier transform that are zero outside the spectrum. Holomorphic continuation of these Fourier transforms then expands the area of these functions being equal to zero and thus allows to make statements on the shape of the spectrum from the outside, including properties of a mass gap. It uses theorems by Pflug, the Edge-of-the-Wedge-Therem, Double Cone Theorem, and the Jost-Lehmann-Dyson-Formula, among others which have already been described and used in \cite{Arm}.

\section{The axiomatic framework}\label{QFT}
In this section, the framework of algebraic quantum field theory will be defined the mentioned axioms.\\

Let $M$ be the (1+3)-dim. Minkowski space\label{minkowskispace} with indefinite scalar product $a b:=a_{0} b_{0}-\sum_{i=1}^{3} a_{i} b_{i}$.

\textbf{Axiom 1} To every bounded open region $O \subset M$ a $C^{*}$-algebra $\mathcal{A}(O)$ is assigned with the \textbf{isotony} property:

\begin{equation*}
O_{1} \subset O_{2} \Longrightarrow \mathcal{A}\left(O_{1}\right) \subset \mathcal{A}\left(O_{2}\right)
\end{equation*}

The measurable observables in $O$ are precisely the self-adjoint elements of $\mathcal{A}(O)$. The norm closure $\mathcal{A}:=\overline{\cup{\mathcal{A}(O) \mid O \subset \mathcal{M} : \textrm{open, bounded}}}$ is called the $C^{\star}$-inductive limit. $\mathcal{A}$ is then again a $C^{*}$-algebra.\\

The \textbf{causality} principle from special relativity is formulated here as\

\textbf{Axiom 2} $O_{1} \subset O_{2}^{\prime}:=\left\{a \in M \mid(a-b)^{2}<0 : \forall b \in O_{2}\right\}$

\begin{equation*}
\Longrightarrow \mathcal{A}\left(O_{1}\right) \subset \mathcal{A}\left(O_{2}\right)^{\prime}:=\left\{x \in \mathcal{B}(\mathcal{H}) \mid[x, y]:=x y-y x=0 :\forall y \in \mathcal{A}\left(O_{2}\right)\right\}
\end{equation*}

Observable quantities from spacelike separated regions should commute.\\

The symmetry group typically acting is the proper orthochronous Poincaré group. However, only translations are required as \textbf{symmetry} here.

\textbf{Axiom 3} The translation group canonically isomorphic to $M$ (which is denoted by $M$ again) acts as a group of automorphisms $\alpha_a$ on $\mathcal{A}$, such that for each bounded open subset $O \subset M$:

\begin{equation*}
\alpha_a \mathcal{A}(O) = \mathcal{A}(O+a) \quad \forall a \in M
\end{equation*}

\begin{definition}
A triple $\{\pi, \mathcal{H}, U\}$ consisting of a non-degenerate representation $\pi$ of $\mathcal{A}$ on the Hilbert space $\mathcal{H}$ together with a continuous unitary representation $U$ of $M$ on $\mathcal{H}$ is called a \textbf{covariant representation} if it respects the automorphisms $\alpha$, i.e., if it satisfies:

\begin{equation*}
U(a) \pi(x) U^{}(a) = \pi(\alpha_{a} x) \quad \forall x \in A, a \in M.
\end{equation*}
\end{definition} 
According to Stone's theorem, there exist self-adjoint operators $P_{0}, \ldots, P_{3}$ such that $U$ can be written as $U(a)=e^{i(P, a)}$; where $(P, a)=(a, P)$ denotes the operator $a_{0} P_{0}-a_{1} P_{1}-a_{2} P_{2}-a_{3} P_{3}$. The spectral decomposition of $P$ looks as follows:

\begin{equation*}
P = \int_{M} p\; dE(p)
\end{equation*}

with the projection-valued spectral measure $E(\cdot)$ on $M$.

This means for $U$:

\begin{equation*}
U(a)=\int\limits_{M} e^{i p a} d E(p) .
\end{equation*}

The following notations are therefore used interchangeably:

\begin{equation*}
\operatorname{spec} U=\operatorname{spec} P=\operatorname{supp}, d E \text {. }
\end{equation*}

\textbf{Axiom 4} The algebra $A$ allows for a faithful covariant representation ${\pi, \mathcal{H}, U}$ in such a way that the spectral measure associated with $U$ satisfies:

\begin{equation*}
\text { supp } d E \subset \overline{V^+} \text { (\textbf{spectrum condition}) }
\end{equation*}

\begin{definition}
A set $\{\mathcal{A}(O), \mathcal{A}, M, \alpha\}$ is called a \textbf{theory of local observables} if Axioms 1-4 are fulfilled.
\end{definition}

By the axioms, $U$ is not uniquely determined. However, in order to introduce the energy-momentum operator through $U$, this uniqueness is required. For this purpose, one defines:

\begin{definition}
Let $U$ be defined by $U(a)=e^{i(P, a)}$ such that Axioms 1-4 are satisfied. Then $U$ is called \textbf{minimal} if for any $U^{\prime}$ with $U^{\prime}(a)=e^{i\left(P^{\prime},a\right)}$, which also satisfies Axioms 1-4, the following holds:

\begin{equation*}
(x, P) \leq\left(x, P^{\prime}\right) \quad \forall x \in \overline{V^{+}} .
\end{equation*}

Here, the $\leq$ sign refers to the order of operators, that is $T_{1} \leq T_{2}$ if and only if $T_{2}-T_{1}$ is positive, i.e., $\left(\Psi,\left(T_{2}-T_{1}\right) \Psi\right) \geq 0 \quad \forall \Psi$ in the domain of $T_{2}-T_{1}$.\
\end{definition}

According to \cite{BoBu}, for any theory of local observables there exists a uniquely determined minimal $U$ that satisfies Axioms 1-4.\par
\hfill \break
The joint spectrum of the energy-momentum operator $P=\left(P_{0}, \ldots, P_{3}\right)$ introduced by $U(a)=e^{i(P, a)}$ is then a Lorentz-invariant set. The proof of this can be found in \cite{Bor3}. Thus, the spectrum condition (Axiom 4) always requires non-negative energy.

For our case, we even demand the strong spectrum condition:

\begin{equation*}
\operatorname{spec} U \subset{0} \cup \overline{V_{\mu}^{+}}
\end{equation*}

i.e., no particles with mass $<\mu$ should occur. A theory in which the strong spectrum condition holds is called a theory with \textbf{mass gap}.

In the closed forward light cone (in $R^{n}, n>2$), there are the following Lorentz-invariant sets:

a) ${0}$

b) $\left\{p \mid p^{2}=0, p_{0} \geq 0\right\}$

c) $\left\{p \mid p^{2}=m^{2}\right.$ for a fixed $\left.m, p_{0}>0\right\}$

d) arbitrary unions of the sets mentioned in a), b), c).\\

If we restrict ourselves to massive particles with charge $Q \neq 0$ (i.e., outside the vacuum sector), possibilities a) and b) are eliminated for spec $U$. Therefore, spec $U$ is a union of hyperboloids.

\section{Holomorphic functions in QFT}\label{HolQFT}
In this section, some specific functions are introduced which will later be expanded to complex-valued domains and used for holomorphic continuation.

However, some notions and terminologies from functional analysis are needed. They can be found, for example, in \cite{RS}.\\

If $\boldsymbol{U}(a)=\int e^{i p a} d E(p)$, then one associates the (orthogonal) spectral projection $E(S)$ to a Borel subset $S \subset M$. If $U(a)$ belongs to a von Neumann algebra $\mathcal{N}$, then $E(S) \in \mathcal{N}$ for every $S$. For $\Psi \in \mathcal{H}$, we denote by supp $\Psi$ the smallest closed set $S \subset$ spec $U$ such that $E(S) \Psi=\Psi$.

\begin{lemma}\label{lem316}
The (continuous) functions $F_{x, \Psi}^{+}$, $F_{x, \Psi}^{-}$ defined by

\begin{equation*}
\begin{aligned}
& F_{x, \Psi}^{+}(a)=\left(\Psi, \pi\left(x^{*}\right) U(a) \pi(x) \Psi\right) \quad \text { and } \
& F_{x, \Psi}^{-}(a)=\left(\Psi, \pi\left(\alpha_{a} x\right) \pi\left(x^{*}\right) U(a) \Psi\right),
\end{aligned}
\end{equation*}

are bounded and can therefore be regarded as distributions in $S^{\prime}(M)$. For their Fourier transforms\label{Fouriertransformierte} $\mathcal{F} F_{x, \Psi}^{+} \equiv \widetilde{F_{x, \Psi}^{+}}$, $\mathcal{F} F_{x, \Psi}^{-} \equiv \widetilde{F_{x, \Psi}^{-}}$, we have:

\begin{enumerate}
\item $\operatorname{supp} \widetilde{F_{x, \Psi}^{+}} \subset \operatorname{spec} U$

\item $\operatorname{supp} \widetilde{F_{x, \Psi}^{-}} \subset 2 \operatorname{supp} \Psi-\operatorname{spec} U$

\end{enumerate}
\end{lemma} 
\textbf{Proof} i) $F_{x, \Psi}^{+}$, $F_{x, \Psi}^{-} \in S^{\prime}(M)$ since

\begin{equation*}
\left|\int_{M}\left(\Psi, \pi\left(x^{*}\right) U(a) \pi(x) \Psi\right) \rho(a) d a\right| \leq\|\Psi\|^{2}\|\pi(x)\|^{2} \int_{M} \rho(a) d a<\infty \quad \forall \rho \in \mathcal{S}(M)
\end{equation*}

The proof for $F_{x, \Psi}^{-}$ is analogous.

ii) Let $S \subset M$ be a Borel set with $S \cap \operatorname{spec} U=\emptyset$. Then $E(S)=0$. For $\rho \in \mathcal{S}(M)$ and $\Phi=\pi(x) \Psi$, we have:

$$(\mathcal{F}(\Phi, U(.) \Phi), \rho)=((\Phi, U(.) \Phi), \tilde{\rho})=\int_{M} \tilde{\rho}(a)(\Phi, U(a) \Phi) d a=\int_{M} \tilde{\rho}(a) \int_{\operatorname{spec} U} e^{i p a}(\Phi, d E(p) \Phi) d a=$$ $$=\int_{\text {spec } U} \int_{M} \tilde{\rho}(a) e^{i p a} d a(\Phi, d E(p) \Phi)=(2 \pi)^{2} \int_{\operatorname{spec} U} \rho(p)(\Phi, d E(p) \Phi)=0, \text{ if supp } \rho \subset S.$$

This shows that supp $\widetilde{F_{x, \psi}^{+}} \subset \operatorname{spec} U$.\par
\hfill \break
The argument for $F_{x, \Psi}^{-}$ is analogous:

\begin{equation*}
\begin{aligned}
& \left(\mathcal{F}\left(F_{x, \psi}^{-}\right), \rho\right)=\int_{M}\left(\Psi, \pi\left(\alpha_{a} x\right) \pi\left(x^{*}\right) U(a) \Psi\right) \tilde{\rho}(a) d a \\
& =\int_{M}\left(\Psi, U(a) \pi(x) U(-a) \pi\left(x^{*}\right) U(a) \Psi\right) \bar{\rho}(a) d a \\
& =\int_{M} \int_{\operatorname{spec} U} \int_{\operatorname{spec} U} \int_{\operatorname{spec} U} e^{i a p} e^{-i a q} e^{i a r}\left(\Psi, d E(p) \pi(x) d E(q) \pi\left(x^{*}\right) d E(r) \Psi\right) \bar{\rho}(a) d a \\
& =\int_{\operatorname{spec} U} \int_{\operatorname{spec} U} \int_{\operatorname{spec} U}\left[\int_{M} e^{i a(p-q+r)} \tilde{\rho}(a) d a\right]\left(\Psi, d E(p) \pi(x) d E(q) \pi\left(x^{*}\right) d E(r) \Psi\right) \\
& =\int_{\text {spec } U} \int_{\operatorname{spec} U} \int_{\operatorname{spec} U} \rho(p-q+r)\left(\pi\left(x^*\right) d E(p) \Psi, d E(q) \pi\left(x^{*}\right) d E(r) \Psi\right) .
\end{aligned}
\end{equation*}

Given $d E(p) \Psi=0$ if $p \notin \operatorname{supp} \Psi$, and $d E(q)=0$ if $q \notin \operatorname{spec} U$, then the value of the integral above is zero if $\operatorname{supp} \rho \cap(\operatorname{supp} \Psi+\operatorname{supp} \Psi-\operatorname{spec} U)=\emptyset$. This was just the claim.\hfill$\blacksquare$\\

\begin{lemma}\label{lem317}
If $S \subset M$ is a Borel set with $E(S) \neq 0$, then the set $\mathcal{K}(S):=\{\pi(x) \Psi \mid x \in \mathcal{A}\left(D_{-t, t}\right)$ for some $t \in V^{+}$, supp $\left.\Psi \subset S\right\}$ is dense in $\mathcal{H}$.
\end{lemma}
\textbf{Proof} For any $x \in \mathcal{A}(O)$ with arbitrary $O \subset M$, it is contained in some $\mathcal{A}\left(D_{-t, t}\right)$ for sufficiently large $t$. The $\pi(x)$ considered in $\mathcal{K}(S)$ are thus dense in $\pi(\mathcal{A})$ with respect to norm convergence, and therefore also with respect to strong convergence. According to von Neumann's density theorem (see e.g.\cite{KR}), $\pi(\mathcal{A})$ is dense in the von Neumann algebra $\pi(\mathcal{A})^{\prime \prime}$ (strong convergence), with $\pi(\mathcal{A})^{\prime \prime}$ being the bicommutant\label{Kommutante} of $\pi(\mathcal{A})$. The $\Psi$ occurring in $\mathcal{K}(S)$ are precisely those located in $E(S) \mathcal{H}$. However, $\overline{\pi(\mathcal{A})^{\prime \prime} E(S) \mathcal{H}}=F \mathcal{H}$, where $F$ denotes the central carrier of $E(S)$ (with $U(a)$ always containing $E(S)$ in $\left.\pi(\mathcal{A})^{\prime \prime}\right)$, see again \cite{KR}. Since $\pi$ is a factor representation, we have because of $E(S) \neq 0\textbf{: } F=\mathbb{1}$. This proves the claim.\hfill$\blacksquare$\\

\section{Techniques of holomorphic continuation}\label{Techniques}
In this section we first summarize several known results on holomorphic continuation as used in QFT. We will then use these results to calculate concrete envelopes of holormorophy for domains that are specific examples of energy-momentum spectrums.
\subsection{Compilation of Some Known Results}
A domain of holomorphy is a connected open set $G \subset \mathbb{C}^{n}$ for which there exists a function $f$ that is holomorphic in $G$, but cannot be holomorphically continued through any boundary point of $G$; that is, for every power series expansion of $f$ around a point $z \in G$ that converges in a poly-cylinder $\Delta(z, R)$ with polyradius $R$, it holds that $\Delta(z, R) \subset G$.

A holomorphy domain $G^{*} \supset G$, into which every holomorphic function in $G$ can be holomorphically continued, is called the simple envelope of holomorphy of $G$. However, not every domain has a simple holomorphic envelope; in general, one obtains a Riemannian domain over $\mathbb{C}^{n}$ as a holomorphic envelope. In both cases, $H(G)$ denotes the holomorphic envelope\label{HolomorphicEnvelope} of $G$.\

\begin{definition} A holomorphic function $f$ on the domain $G$ is said to have \textbf{polynomial growth} if there exist $N \in \mathbb{N}$ and $c>0$ such that for all $z \in G$,
\end{definition}
$$ |f(z)| \leq c\left(\Delta_{G}(z)\right)^{-N}, $$

where $\Delta_{G}$ is defined for $z \in G$ as

\begin{equation*} \Delta_{G}(z):=\min \left\{\operatorname{dist}(z, \partial G),\left(1+\|z\|^{2}\right)^{-1 / 2}\right\}. \end{equation*}

With these notations, the following holds:
\begin{theorem}[Pflug]
Let $G \subset G^{\prime}$ be domains with $H\left(G^{\prime}\right) \subset \mathbb{C}^{n}$. If every holomorphic function on $G$ with bounded growth can be holomorphically continued to $G^{\prime}$, then every holomorphic function on $G$ can be holomorphically continued to $G^{\prime}$ (and hence also to $H\left(G^{\prime}\right)$).
\end{theorem}
\textbf{Proof} According to \cite{Pf2}, the statement holds for the class of functions defined by
$$
|f(z)| \leq c\left(\widetilde{\Delta_{G}}(z)\right)^{-N}
$$
where $\widetilde{\Delta_{G}}(z)=\left(1+\|z\|^{2}\right)^{-1 / 2} \min \{1, \operatorname{dist}(z, \partial G)\}$. However, the class of functions described by $\Delta_{G}$ introduced in \cite{Pf1} coincides with the class described by $\widetilde{\Delta_{G}}$, as it always holds that:

\begin{equation*} \left(\Delta_{G}(z)\right)^{2} \leq \widetilde{\Delta_{G}}(z) \leq \Delta_{G}(z). \end{equation*}
\hfill$\blacksquare$

\begin{definition} A 2-dimensional \textbf{analytic surface} is a set $F \subset \mathbb{C}^{n}$ for which, for every point $z_{0} \in F$, there exists a domain $U \subset \mathbb{C}$ and a vector-valued function $h_{z_{0}}: U \longrightarrow \mathbb{C}^{n}$ such that:

\begin{enumerate}
\item $z_{0}=h_{z_{0}}\left(\lambda_{0}\right)$ for some $\lambda_{0} \in U$
\item $h_{z_{0}}$ is holomorphic in $U$
\item $\left\{z=h_{z_{0}}(\lambda) \mid \lambda \in U\right\}$ represents $F$ near $z_{0}$
\item The vector-valued function $\frac{d}{d \lambda} h_{z_{0}}$ does not vanish anywhere in $U$.
\end{enumerate}
\end{definition}

By the implicit function theorem, such an analytic surface $F \subset \mathbb{C}^{n}$ can always be regarded as a complex submanifold of $\mathbb{C}^{n}$.

The concept of an analytic surface is now used to explicitly specify a domain that is larger than the original domain and still lies entirely within its envelope of holomorphy.

\begin{theorem}[Weak continuity theorem]\label{WCT} Let $\left(G_{\alpha}\right){\alpha \in \mathbf{N}}$ be a sequence of connected sets that are open in 2-dimensional analytic surfaces $F{\alpha}$, and let $\overline{G_{\alpha}} \subset F_{\alpha}$ always hold. Suppose $G \subset \mathbb{C}^{n}$ is a domain (with $\left.\partial G_{\alpha}=\overline{G_{\alpha}} \backslash G_{\alpha}\right)$ such that:

\begin{enumerate}
\item $G_{\circ} \subset \subset G$
\item $\lim \limits_{\alpha \rightarrow \infty} G_{\alpha}=: S_{0}$
\item $\lim \limits_{\alpha \rightarrow \infty} \partial G_{\alpha}=: T_{0} \subset \subset G$
\item $S_{0}$ is bounded
\end{enumerate}

Then, $S_{0} \subset H(G)$.
\end{theorem}

The convergence of $\left(G_{\alpha}\right)$ and $\left(\partial G_{\alpha}\right)$ is to be understood as follows:

One says that the sequence of sets $\left(A_{k}\right), A_{k} \subset \mathbb{C}^{n}$, converges to the set $A \subset \mathbb{C}^{n}$ ($\lim {k \rightarrow \infty} A{k}=A$) if $A$ consists precisely of the limits of all convergent sequences $\left(a_{k}\right)$ in $\mathbb{C}^{n}$ with $a_{k} \in A_{k}$.

The proof of the weak continuity theorem can be found, for example, in \cite{Vl3} and \cite{Bre}. What is essential to it is that 2-dimensional analytic surfaces satisfy the maximum principle with respect to the moduli of holomorphic functions. In other words, for any holomorphic function $f$ on the bounded set $\overline{G_{\alpha}}$ holds:

\begin{equation*}
\sup _{z \in \overline{G_{\alpha}}}|f(z)|=\sup _{z \in \partial G_{\alpha}}|f(z)|
\end{equation*}

\begin{definition} In the following, the scalar product of two vectors in $\mathbb{R}^{n}$ always refers to the indefinite Minkowski product: $a b=a_{0} b_{0}-a_{1} b_{1}-\ldots-a_{n-1} b_{n-1}$. Also, for $a, b \in \mathbb{C}^{n}$, the Minkowski scalar product with this calculation rule should always be understood as $ab$.

The \textbf{forward light cone}\label{forwardcone} is the set $V^{+}:=\left\{x \in \mathbb{R}^{n} \mid x^{2}>0, x_{0}>0\right\}$; $V^{-}:=-V^{+}$ denotes the \textbf{backward light cone}\label{backwardcone}.

The \textbf{forward tube}\label{forwardtube} is the set $T^{+}:=\left\{z=x+i y \in \mathbb{C}^{n} \mid y \in V^{+}\right\}= \mathbb{R}^{n}+i V^{+}$; $T^{-}:=-T^{+}=\mathbb{R}^{n}+i V^{-}$ is the \textbf{backward tube}\label{backwardtube}.
\end{definition}

The transition from real to complex functions is often carried out in quantum field theory by the following statement:

\begin{theorem}\label{th26} Let $f^{+}, f^{-} \in \mathcal{S}^{\prime}\left(\mathbb{R}^{n}\right)$ be tempered distributions, and let $a, b \in \mathbb{R}^{n}$ with $\operatorname{supp} f^{+} \subset a+V^{+}, \operatorname{supp} f^{-} \subset b+V^{-}$.

Then the Fourier transforms $\mathcal{F} f^{+}, \mathcal{F} f^{-}$ of $f^{+}$ and $f^{-}$ are boundary values in the distributive sense of functions that are holomorphic in $T^{+}$ and $T^{-}$, respectively; that is, there exist holomorphic functions $G^{+}, G^{-}$ in $T^{+}$ and $T^{-}$, respectively, such that for all $\phi \in \mathcal{S}\left(R^{n}\right)$:

$$
\left(\mathcal{F} f^{+}, \phi\right)=\lim _{y \rightarrow 0, y \in V^{+}} \int G^{+}(x+i y) \phi(x) d x
$$

independently of the sequence chosen for $y \rightarrow 0$ (analogous for $\mathcal{F} f^{-}$).\
\end{theorem}
The proof can be found, for example, in \cite{Bog} or \cite{Bor5}.\par
\hfill \break
Thus, in certain applications, one deals with functions that are holomorphic in $T^{+}$ or $T^{-}$. The following theorem deals with a situation in which limits of such functions coincide in certain real regions:

\begin{theorem}[Edge-of-the-Wedge-Theorem]\label{EOTW} Let $f^{+}, f^{-}$ be functions that are holomorphic in $\mathrm{T}^{+}$ and $T^{-}$, respectively, and let there be a region $B \subset \mathbb{R}^{n}$ for which $f^{+}$ and $f^{-}$ have matching boundary values in the distributive sense. Then there is a function $f$ and a complex neighborhood $\tilde{B}$\label{tildeB} of $B$, such that $f$ is holomorphic on $T^{+} \cup T^{-} \cup \tilde{B}$ and $f \mid T^{+}=f^{+}$ as well as $f \mid T^{-}=f^{-}$.
\end{theorem}

\begin{remark}\label{rem28}
One can provide further information on the size and shape of the neighborhood $\tilde{B}$; we set:

$$
\tilde{B}:=\bigcup_{x \in B}\left\{z \mid\|z-x\|<\frac{1}{32} \operatorname{dist}(x, \partial B)\right\} .
$$

In this situation, the following holds for this $\tilde{B}$:

\begin{enumerate}
\item $\tilde{B} \cap \mathbb{R}^{n}=B$, no further real points are added.

\item $x \notin B \Rightarrow$ there is no $y \in \mathbb{R}^{n}$ with $x+i y \in \tilde{B}$.

\item $\operatorname{dist}(x, \partial \tilde{B})$ in $\mathbb{C}^{n}$ is proportional to $\operatorname{dist}(x, \partial B)$ in $\mathbb{R}^{n}$ for real $x$.

\end{enumerate}
\end{remark} 
The proof of the Edge-of-the-Wedge theorem and this remark can be found in \cite{Vl3}.\par
\hfill \break
For the following, we need:

\begin{definition}
 Let $x, y \in \mathbb{R}^{n}$ be two points with $y \in x+V^{+}$. Then,

$$
D_{x, y}:=\left(x+V^{+}\right) \cap\left(y+V^{-}\right)
$$

denotes the \textbf{double cone}\label{double cone} spanned by $x$ and $y$. 
\end{definition} 

For the shape of the real coincidence region $B$ from the Edge-of-the-Wedge Theorem, we have the following:

\begin{theorem}[Double Cone Theorem]\label{DCT} Let $x, y \in B$ be two points that can be connected by a timelike curve (i.e., its tangent in every point is timelike) entirely within the interior of $B$, and let $y \in x+V^{+}$. Then, we have:

$$
D_{x, y} \subset H\left(T^{+} \cup T^{-} \cup \tilde{B}\right) \cap \mathbb{R}^{n} \text {. }
$$
\end{theorem}

In applications, we always deal with light cone convex regions, such as double cones.\par
\hfill \break
The proof of the Double Cone Theorem can be found, among others, in \cite{Vl1}. It is proven there using the weak continuity theorem. Another proof can be found in \cite{Bor1}, following an idea from \cite{BM}. There, the Cauchy integral formula is used for the holomorphic continuation.

Both methods will be used here to prove Theorem \ref{th217}.\par
\hfill \break
For certain real coincidence regions $G$, it is even possible to explicitly determine the holomorphic envelope of $T^{+} \cup T^{-} \cup \tilde{G}$. The following preparations are used for this purpose.

Let $M \subset \mathbb{R}^{r}$ be an open set. The hyperboloid

$$
\left(x-x^{\prime}\right)^{2}=\lambda^{2}, x^{\prime} \in \mathbb{R}^{n}, \lambda \in \mathbb{R}^{+}
$$

is said to be admissible to $M$\label{admissibleHyp} if

$$
M \cap\left\{x \mid\left(x-x^{\prime}\right)^{2} \geq \lambda^{2}\right\}=\emptyset
$$

i.e., if $M$ lies between the branches of the hyperboloid. The set of parameters $\left(x^{\prime}, \lambda\right)$ corresponding to admissible hyperboloids is denoted by $N(M)$:

$$
N(M):=\left\{\left(x^{\prime}, \lambda\right) \in \mathbb{R}^{n} \times \mathbb{R}^{+} \mid\left(x-x^{\prime}\right)^{2}<\lambda^{2} \; \forall x \in M\right\}
$$

It is possible that $N(M)=\emptyset$. For such sets $M$, we define the set $N_{\infty}(M)$\label{admissibleinfinity} of parameters corresponding to admissible hyperplanes:

$$
N_{\infty}(M):=\left\{\left(x^{\prime}, a\right) \in \mathbb{R}^{n} \times\left(V^{+} \cup V^{-}\right)^{-} \mid a\left(x-x^{\prime}\right)<0 \; \forall x \in M\right\}
$$

These terms can be used to formulate:

\begin{theorem}[Jost-Lehmann-Dyson formula]\label{JLD} Let $G \subset \mathbb{R}^{n}$ be a connected open set bounded by two space-like hyperplanes, that is,

$$
G=\left\{x=\left(x_{0}, \tilde{x}\right) \mid f(\tilde{x})<x_{0}<g(\tilde{x})\right\}
$$

with functions $f, g$ satisfying:

$\left|f\left(\tilde{x}_{1}\right)-f\left(\tilde{x}_{2}\right)\right| \leq\left|\tilde{x}_{1}-\tilde{x}_{2}\right|$ and $\left|g\left(\tilde{x}_{1}\right)-g\left(\tilde{x}_{2}\right)\right| \leq\left|\tilde{x}_{1}-\tilde{x}_{2}\right| \quad \forall \tilde{x}_{1}, \tilde{x}_{2} \in \mathbb{R}^{n-1}$.

Furthermore, assume that $N(G) \neq \emptyset$. Then, for the holomorphic envelope of $T^{+} \cup T^{-} \cup \tilde{G}$, we have:

$$
H(T^{+} \cup T^{-} \cup \tilde{G}) = \mathbb{C}^{n} \setminus \overline{\bigcup_{\left(x^{\prime},\lambda\right) \in N(G)} \left\{z \mid \left(z-x^{\prime}\right)^{2} = \lambda^{2}\right\}}.
$$
\end{theorem}
The proof can be found in \cite{BMS}.

\begin{remark}\label{rem212} Suppose that for the domain $B=\bigcup B_{i}$ (where $B_{i}$ are the connected components), we always have $\left(x_{i}-x_{j}\right)^{2}<0$, if $x_{i} \in B_{i}, x_{j} \in B_{j}, i \neq j$, i.e., all the connected components are space-like to each other, and each $B_{i}$ satisfies the conditions of Theorem \ref{JLD}. Then, the holomorphic envelope of $\tilde{B} \cup T^{+} \cup T^{-}$ is obtained using the same procedure as in Theorem \ref{JLD}.
\end{remark}
A proof of this remark can be found in \cite{Vl2}.
\\
In the case where $N(G)=\emptyset$, we obtain:

\begin{corollary}\label{cor213}
Let $G \subset \mathbb{R}^{n}$ be an open set with $N(G)=\emptyset$ and suppose that

$$
G=\left(G+V^{+}\right) \cap\left(G+V^{-}\right)
$$

Then:

$$
H(T^{+} \cup T^{-} \cup \tilde{G}) = \mathbb{C}^n \backslash \overline{\bigcup_{\left(x^{\prime}, a\right) \in N_{\infty}\left(G\right)}\left\{z \mid a\left(z-x^{\prime}\right)=0\right\}}.
$$
\end{corollary}

\textbf{Proof} The regions $G_{\alpha}=G \cap\left\{x|| x_{0} \mid<\alpha\right\}$ satisfy the conditions of Theorem \ref{JLD} for $\alpha>0$ with $f=-\alpha,\; g=\alpha$. With $K\left(G_{\alpha}\right):=H\left(T^{+} \cup T^{-} \cup \tilde{G}_{\alpha}\right)$ and\\ $K(G):=\mathbb{C}^{n} \setminus \overline{\bigcup\limits{\left(x^{\prime}, a\right) \in N_{\infty}(G)}\left\{z \mid a\left(z-x^{\prime}\right)=0\right\}}$, we have, according to \cite{Vl3} (Section 33.1):

\begin{equation*}
K(G)=\bigcup_{\alpha=1}^{\infty} K\left(G_{\alpha}\right)
\end{equation*}

For each $\alpha$, $K\left(G_{\alpha}\right)$ is a domain of holomorphy, and therefore, by the Behnke-Stein theorem (see \cite{Vl3}), $K(G)$ is also a domain of holomorphy, and it holds that $H\left(T^{+} \cup T^{-} \cup \tilde{G}\right)=K(G)$.\hfill$\blacksquare$\
\subsection{Application to specific cases}\label{sec22}

The results of the first section shall now be applied to specific domains in $\mathbb{C}^n$, $n\geq 2$. For $\mu\geq 0$, let $V_\mu^{+}$ be the following set: \label{Vmuplus}
\begin{equation*}
V_\mu^{+}:=\left\{x \in \mathbb{R}^n \mid x^2>\mu^2, x_0>0\right\}
\end{equation*}
In the case $\mu=0$, we have $V_0^{+}=V^{+}$.\

To prepare for this, it is first necessary to make some observations in the case $n=2$. Let $\hat{y}$ denote the timelike vector uniquely determined by a spacelike vector $y \in \mathbb{R}^2$ through
\begin{equation*}
\hat{y}^2=1, y \hat{y}=0, \hat{y} \in V^{+}
\end{equation*}
given by:
\begin{equation*}
\hat{y}=\frac{\operatorname{sgn} y_1}{\sqrt{-y^2}}\left(y_1, y_0\right).
\end{equation*}
\begin{theorem}\label{th214} In $\mathbb{C}^2$, for $\mu \geq 0$:
\begin{equation*}
\begin{aligned}
H\left(\widetilde{V_\mu^{+}} \cup T^{+} \cup T^{-}\right)= \quad & T^{+} \cup T^{-} \cup\left\{z=x+i y \in \mathbb{C}^2 \mid y^2<0 \text { and } x \hat{y}>\mu\right\} \cup \\
& \cup\left\{z \mid y^2=0, y \neq 0, x_0>x_1 \operatorname{sgn} y_0 \operatorname{sgn} y_1\right\} \cup\left\{z \mid y=0, x \in V_\mu^{+}\right\} .
\end{aligned}
\end{equation*}
In particular, for $\mu=0$, the case of $V^{+}$ as a coincidence domain is included.
\end{theorem}

\textbf{Proof} For $V_\mu^{+}$, the assumptions of Corollary \ref{cor213} are satisfied. Therefore, the holomorphic envelope can be calculated using the procedure there. The condition $a\left(x-x^{\prime}\right)<0 \quad \forall x \in V_\mu^{+}$ means that the line defined by $a\left(x-x^{\prime}\right)=0$ does not intersect the region $V_\mu^{+}$. It can be shown that:
\begin{equation*}
N_{\infty}\left(V_\mu^{+}\right)=\left\{\left(x^{\prime}, a\right) \mid a \in \overline{V^{-}}, x^{\prime} a<-\mu \sqrt{a^2}\right\}
\end{equation*}
According to Corollary \ref{cor213}, we have:
\begin{equation*}
H\left(\widetilde{V_\mu^{+}} \cup T^{+} \cup T^{-}\right)=\mathbb{C}^n \backslash \overline{\bigcup_{\left(x^{\prime}, a\right) \in N_{\infty}\left(V_\mu^{+}\right)}\left\{z \mid \;a\left(z-x^{\prime}\right)=0\right\}}.
\end{equation*}

Here, $a\left(z-x^{\prime}\right)$ means that, for spacelike $y$, we have:\par
(1) $a\left(x-x^{\prime}\right)=0$\par
(2) $a y=0 \Rightarrow a=-\hat{y}$\par
\hfill \break
Using (1), we have: $\hat{y}\left(x-x^{\prime}\right)=0$ or $x-x^{\prime}=\alpha y$, that is, $x=\alpha y+x^{\prime}$.\\
In $N_{\infty}\left(V_\mu^{+}\right)$, all $x^{\prime}$ lying below the line $\{\mu \hat{y}+c y \mid c \in \mathbb{R}\}$ are included. Therefore, in the complement of $H\left(\widetilde{V_\mu^{+}} \cup T^{+} \cup T^{-}\right)$, for fixed $y$, all $x$ lying below $\{\alpha y+\mu \hat{y}+c y\}= \left\{\mu \hat{y}+c^{\prime} y\right\}$ are included. This means that for $z \in H\left(\widetilde{V_i^{+}} \cup T^{+} \cup T^{-}\right)$, we have $x \hat{y}>\mu$.\\
For lightlike $y \neq 0$, we need to take the closure of the points $y$ with $y^2<0$ that do not lie in the holomorphic envelope. This means: $x \hat{y} \leq m \Leftrightarrow x_0\left|y_1\right|-x_1 y_0$ sgn $y_1 \leq m \sqrt{-y^2}$. Taking the limit $y_1 \longrightarrow y_0 \neq 0$ yields $x_0-x_1$ sgn $y_0$ sgn $y_1 \leq 0$, which proves the claim.\\
For $y=0$, the assertion follows directly from the definition of $N_\infty$. $\blacksquare$

\begin{corollary}\label{cor215}
Let $G\subset\mathbb{R}^2$ be a domain such that $G+V^+=G$. Then for any line $g$ given by $g(t)=a+ty$ where $t\in\mathbb{R}$ and $a,y\in\mathbb{R}^2$ with $y\neq 0$ that intersects $G$, we have:
\begin{equation*}
g(t+i\tau)\in H(\tilde{G}\cup T^+\cup T^-)\quad\forall\tau\neq 0
\end{equation*}
\end{corollary} 
\textbf{Proof} If $g$ intersects $G$, then there exists $b\in G$ such that $g$ intersects $b+V^+\subset G$. But the holomorphic envelope of $(b+\widetilde{V^+})\cup T^+\cup T^-$ is known.

Suppose $g(t+i\tau)$ is a point with $\tau\neq 0$. Then $\tau y=\operatorname{Im}(g(t+i\tau))\neq 0$. The assertion is clear if $y^2\geq 0$.

If $y^2<0$, then we can translate the real coordinate system by $-b$ without affecting the imaginary part. Thus, $b$ is moved to the origin and $g$ becomes $g':=g-b$. Let $x:=\operatorname{Re}(g'(t+i\tau))=\operatorname{Re}(g(t+i\tau))-b$. Then there exists $\tilde{x}\in g'\cap V^+$ such that $x=\tilde{x}+\lambda y$ for some $\lambda\in\mathbb{R}$. Hence, $x\hat{y}=\tilde{x}\hat{y}>0$ since $\tilde{x}\in V^+$. By Theorem \ref{th214}, we have $x+iy\in H(\widetilde{V^+}\cup T^+\cup T^-)$. Therefore, $g(t+i\tau)\in H((b+\widetilde{V^+})\cup T^+\cup T^-)\subset H(\tilde{G}\cup T^+\cup T^-)$. $\blacksquare$

\begin{definition}
Given a set $M\subset\mathbb{R}^n$, we define $M'$\label{spacelikeSet} as the set of \textbf{spacelike points} with respect to every element of $M$:
\begin{equation*}
M':={x\in\mathbb{R}^n\mid (x-y)^2<0\ \forall y\in M}
\end{equation*}
We denote $R:={0}'$ as the set of spacelike points.\label{spacelike}
\end{definition} 

\begin{theorem}\label{th217} Any function that is holomorphic in $\widetilde{V_\mu^{+}} \cup \widetilde{D_{0, s}^{\prime}} \cup T^{+} \cup T^{-},\, s \in V^{+}$, can be holomorphically extended to $R$.
\end{theorem}
\textbf{Proof:} Due to the rotational symmetry of the original domain, it suffices to consider the case $n=2$. Let us choose a point $p \in \partial D_{0,s}^{\prime} \cap R$ (without loss of generality, let $p_1>0$) and select a line $g$ with slope magnitude $<1$ passing through $p$ and intersecting $V_\mu^{+}$. We also consider the right branch of a hyperbola whose apex is at $q:=g \cap d$ with $d:=\left\{x \mid x_0=x_1\right\}$ and whose asymptotes are $g$ and $d$. It is clear that by choosing the hyperbola parameter suitably, we can make it pass through any point of the triangle $\Delta_{p q r}$ with $r:=d \cap{s+\lambda(1,-1), \lambda \in \mathbb{R}}$. This hyperbola shall be parameterized by $K(t), t \in \mathbb{R}$. For a compact $t$-interval, $K(t)$ lies in $\Delta_{p q r}$, and the tangent to $K(t)$ always intersects the region $V_\mu^{+}$.

\begin{figure}[h]
\centering
\includegraphics{"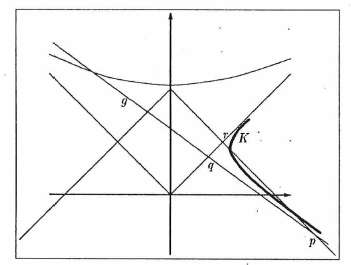"}
\caption{Positioning of the curve $K$ in the proof of Theorem \ref{th217}}
\label{Abbildung 1}
\end{figure}

Now consider the family of curves $K_\alpha$, obtained by shifting $K$ to the right by $\alpha$ parallel to the direction of the line $d$. For $\alpha>\max_{t \in \mathbf{R}}$ (dist $\left(K(t), D_{0, s}^{\prime}\right)$, we have $K_\alpha \subset D_{0,s}^{\prime}$, and $K_0=K$. By replacing $t$ with $t+i \tau$ in the complex plane, we obtain from \ref{cor215} that $h(t+i \tau, \alpha) \in H\left(\widetilde{V_\mu^{+}} \cup T^{+} \cup T^{-}\right)$ for small $\tau \neq 0$, since the imaginary part points in the direction of the derivative with respect to $\tau$, which intersects $V_\mu^{+}$ according to the construction. Moreover, $h(t, \alpha) \in D_{0, s}^{\prime} \subset H\left(\widetilde{V_{\mu}^{+}} \cup \widetilde{D_{0, s}^{\prime}} \cup T^{+} \cup T^{-}\right)$ outside a compact $t$-interval.

Suppose now that $f$ is holomorphic in $\widetilde{V_{\mu}^{+}} \cup \widetilde{D_{0, s}^{\prime}} \cup T^{+} \cup T^{-}$. Then we define
\begin{equation*}
\phi(t+i \tau, \alpha)=\frac{1}{2 \pi i} \oint_W \frac{f(h(\rho, \alpha))}{\rho-(t+i \tau)} d \rho
\end{equation*}

is a holomorphic function inside of $W$. Here, $W$ is a closed curve in the $(t+i \tau)$ plane that contains all values of $t$ for which $K(t) \notin D_{0, s}^{\prime}$ in its interior. Furthermore, $W$ is smooth, and it holds that $h(w, \alpha) \in H\left(\widetilde{V_\mu^{+}} \cup T^{+} \cup T^{-}\right)$ if $w \in W$ and $Im\, w \neq 0$. Since $\phi$ coincides with $f$ for $\alpha > \max_{t \in \mathbf{R}}\left(\operatorname{dist}\left(K(t), D_{0, s}^{\prime}\right)\right)$, $\phi$ is a holomorphic extension of $f$ to $\bigcup_{\alpha \geq 0} K_\alpha$.

Since one can cover $R$ entirely with hyperbolas of the type described above, it follows that $R \subset H\left(\widetilde{V_\mu^{+}} \cup \widetilde{D_{0, s}^{\prime}} \cup T^{+} \cup T^{-}\right)$.\hfill$\blacksquare$\

Alternatively, one can prove the previous theorem \ref{th217} using the weak continuity theorem:

Consider a family of curves $C_\alpha$, $0 \leq \alpha \leq \alpha_0$, with the following properties:
\begin{enumerate}
\item $C_\alpha \subset D_{0,s}^{\prime}$ for $0 < \alpha$
\item $C_0 \cap \partial D_{0, s}^{\prime} \cap R={X}$, $C_0 \backslash{X} \subset D_{0, s}^{\prime}$
\item $C_a$ are real-analytic curves given by the equations
\begin{equation*}
x_j=x_{j, \alpha}(\xi),\quad j=0,1, \quad 0 \leq \alpha \leq \alpha_0, \quad 0 \leq \xi \leq 1
\end{equation*}
\item Every tangent to $C_\alpha$ lies on a line that intersects $V_\mu^{+}$.
\end{enumerate}

\begin{figure}[h]
 \centering
 \includegraphics{"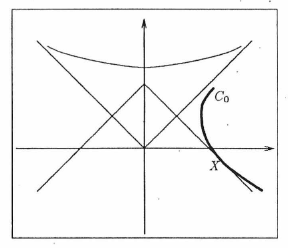"}
 \caption{Position of the curve $C_0$}
 \label{Figure 2}
\end{figure}

The existence of such a family of curves is clear. One can, for example, take the right branch of a hyperbola (with non-lightlike asymptotes) passing through $X$ and having the tangent $\{x \mid x=s+\lambda(-1,1), \lambda \in \mathbb{R}\}$ at $X$. Then $C_0$ is the part of the hyperbola whose tangents intersect $V_\mu^{+}$, and $C_\alpha$ is the curve obtained by shifting the hyperbola by $\alpha$ into the interior of $D_{0, s}^{\prime}$. We now analytically continue the $C_\alpha$ for complex values of $\lambda=\xi+i \eta$. Then $x_{j, \alpha}$ is holomorphic in $A_\delta=\{|\eta|<\delta, 0 \leq \xi \leq 1\}$ for some small $\delta$. In this way, we have constructed 2-dimensional analytic surfaces $F_\alpha$ containing the curves $C_\alpha$, given by $z_j=x_{j, \alpha}(\lambda)$, $j=0,1$, $0 \leq \alpha \leq \alpha_0$, $\lambda \in A_\delta$.

Let $t(\xi)$ be the tangent to $C_\alpha$ at the point characterized by $\xi$. Then, by Taylor expanding $x_{j, \alpha}(\lambda)$ about the point $\xi$, we have $\operatorname{Im},z_j=\eta, t(\xi)j+O(\eta^2)$, meaning that for small $\eta$, the imaginary part of $F_\alpha$ points in the direction of the tangent to $C_\alpha$. Since a spacelike tangent still intersects $V_\mu^{+}$, all points of $F_\alpha$ for $\eta \neq 0$ still belong to $H(\widetilde{V_\mu^{+}} \cup T^{+} \cup T^{-})$ (as argued in the proof above).

However, since $C_\alpha \subset D_{0, s}^{\prime}$ for $0<\alpha$, and $C_0 \backslash{X} \subset D_{0, s}^{\prime}$, it follows from the weak continuity theorem \ref{WCT} that:
\begin{equation*}
C_0 \subset H\left(\widetilde{V_\mu^{+}} \cup \widetilde{D_{0, s}^{\prime}} \cup T^{+} \cup T^{-}\right) \text {. }
\end{equation*}

However, there is also a neighborhood $U(X)$ in $H\left(\widetilde{V_\mu^{+}} \cup \widetilde{D_{0, s}^{\prime}} \cup T^{+} \cup T^{-}\right)$ for $X$. Using the double cone theorem \ref{DCT}, we see that the entire strip $\left(U(X)+V^{+}\right) \cap\left(D_{0, s}^{\prime}+V^{-}\right)$ also lies in $H\left(\widetilde{V_\mu^{+}} \cup \widetilde{D_{0, s}^{\prime}} \cup\right.$ $T^{+} \cup T^{-}$).

If we carry out the procedure indicated here with further points from $D_{0, s}^{\prime} \cap R$, we obtain the union of some (possibly infinitely many) strips that still lie in $H\left(\widetilde{V_\mu^{+}} \cup \widetilde{D_{0, s}^{\prime}} \cup T^{+} \cup T^{-}\right)$. Of course, the procedure can also be carried out for points on the boundary of such strips.

It is conceivable that this procedure terminates at a boundary curve that has $D_{0, s}^{\prime} \cap R$ or a parallel as an asymptote. For this case, consider a line with a space-like direction vector that intersects $V_\mu^{+}$ and lies in a compact piece of $R \backslash F$ ($F$ denotes the set of points in $R$ where a continuation has already been achieved). Now move this line in the direction of the line $\left\{x_0=x_1\right\}$ until the compact piece in $R \backslash F$ consists only of boundary points of $F$. Through these boundary points (possibly only one), we can continue using the method given above.

Overall, we obtain:

\begin{equation*}
R \subset H\left(\widetilde{V_\mu^{+}} \cup \widetilde{D_{0, s}^{\prime}} \cup T^{+} \cup T^{-}\right)
\end{equation*}
\hfill$\blacksquare$\par
\hfill \break
The two proof methods presented here correspond to those in the proofs of the double cone theorem \ref{DCT} in \cite{Bor1} and \cite{Vl1} (see also the explanations following this theorem).

\begin{remark}\label{rem218} With the same reasoning as in the proofs of the preceding theorem, one can holomorphically continue from $\widetilde{V_\mu^{+}} \cup \widetilde{D_{a, b}} \cup T^{+} \cup T^{-}$ to $\widehat{D_{a, b}}$ in $\mathbb{C}^2$. $\widehat{D_{a, b}}$ is obtained from $D_{a, b}$ for $a, b \notin \overline{V^-}$ by the following construction:\
The lines that touch $V_\mu^{+}$ and pass through $a$ and $b$ are denoted by $g_a$ and $g_b$, respectively. The segment from $a$ to the point of contact of $g_a$ with $V_\mu^{+}$ is denoted by $s_a$, and the half-line on $g_b$ that moves away from $V_\mu^{+}$ starting from $b$ is denoted by $s_b$. Then:
\begin{equation*}
\widehat{D_{a, b}}=\left(\left(s_a+V^{+}\right) \cap\left(D_{a, b}+V^{-}\right)\right) \cup\left(\left(s_b+V^{-}\right) \cap\left(D_{a, b}+V^{+}\right)\right)
\end{equation*}
If $a \notin \overline{V^{-}}$, $b \in \overline{V^-}$, then:
\begin{equation*}
\widehat{D_{a, b}}=\left(s_a+V^{+}\right) \cap\left(D_{a, b}+V^{-}\right)
\end{equation*}
For $a \in \overline{V^-}$, $b \notin \overline{V^-}$, we obtain:
\begin{equation*}
\widehat{D_{a, b}}=\left(s_b+V^{-}\right) \cap\left(D_{a, b}+V^{+}\right)
\end{equation*}

However, if $D_{a, b} \subset V^{+}$, then $\widehat{D_{a, b}}=D_{a, b}$.\
\end{remark}
\begin{definition} The mapping $\phi: \mathbb{C}^2 \backslash\left\{z^2=0\right\} \longrightarrow \mathbb{C}^2 \backslash\left\{z^2=0\right\}$, defined by $\phi(z)=\frac{-z}{z^2}$, is called the transformation of reciprocal radii.
\end{definition}
\begin{lemma}\label{lem220}
$\phi$ has the following properties:
\begin{enumerate}
\item $\phi \circ \phi(z)=z \quad \forall z \in \mathbb{C}^2 \backslash\left\{z^2=0\right\}$. Since $\phi$ is holomorphic, it is also biholomorphic and thus $\phi^{-1}=\phi$
\item $\phi\left(T^{+}\right)=T^{+}, \phi\left(T^{-}\right)=T^{-}$
\item $\phi\left(V^{+}\right)=V^{-}$
\item $\phi(R)=R$
\item $\phi\left(D_{\left(-\frac{1}{m}, 0\right), 0}\right)=(m, 0)+V^{+}$\
\end{enumerate}
\end{lemma} 
\textbf{Proof} 1)
\begin{equation*}
\phi \circ \phi(z)=\phi\left(-\frac{z}{z^2}\right)=\frac{-\frac{-z}{z^2}}{\left(\frac{-z}{z^2}\right)^2}=\frac{z}{z^2} \frac{z^4}{z^2}=z .
\end{equation*}

2) Let $z \in T^{+}$. Then
\begin{equation*}
\begin{aligned}
\operatorname{Im} \phi(z) & =\operatorname{Im} \left(\frac{-z}{z^2}\right)=\operatorname{Im} \frac{-x-i y}{x^2-y^2+2 i x y}=\operatorname{Im} \frac{(-x-i y)\left(x^2-y^2-2 i x y\right)}{\left(x^2-y^2\right)^2+4(x y)^2} \\
& =\frac{1}{\left(x^2-y^2\right)^2+4(x y)^2}\left(2 x(x y)-y\left(x^2-y^2\right)\right)
\end{aligned}
\end{equation*}
and therefore
\begin{equation*}
\begin{split}
(\operatorname{Im} \phi(z))^2 & =\frac{1}{()^2} \left[4 x^2(x y)^2+y^2\left(x^2-y^2\right)^2-4(x y)(x y)\left(x^2-y^2\right) \right]  \\ 
& =\frac{1}{()^2}\left[y^2\left(x^2-y^2\right)^2+4 y^2(x y)^2\right]>0, \text{ since } y^2>0 \text{ and } z^2 \neq 0.
\end{split}
\end{equation*} 
3) Let $x \in V^{+}$. Then
\begin{equation*}
(\phi(x))^2=\left(\frac{-x_0}{x^2}\right)^2-\left(\frac{-x_1}{x^2}\right)^2=\frac{1}{x^2}>0 \text{ and }-\frac{x_0}{x^2}<0,
\end{equation*}
so $\phi(x) \in V^{-}$.\

4) Let $x^2<0$. Then we have:
\begin{equation*}
(\phi(x))^2=\left(-\frac{x}{x^2}\right)^2=\frac{x^2}{x^2 x^2}=\frac{1}{x^2}<0 .
\end{equation*}

5) It holds that:
\begin{equation*}
\begin{aligned}
\frac{-x}{x^2} \in(m, 0)+V^{+} & \Leftrightarrow \frac{-x_0}{x^2}-m>\left|\frac{x_1}{x^2}\right| \Leftrightarrow \frac{-x_0}{x^2}-m>\frac{\left|x_1\right|}{x^2} \text { and } x^2>0 \\
& \Leftrightarrow 0<x^2<-\frac{1}{m}\left(x_0+\left|x_1\right|\right) \Leftrightarrow\left(x_0+\frac{1}{2 m}\right)^2<\left(\left|x_1\right|-\frac{1}{2 m}\right)^2 \\
& \Leftrightarrow\left|x_0+\frac{1}{2 m}\right|<|| x_1\left|-\frac{1}{2 m}\right| \Leftrightarrow-\frac{1}{m}+\left|x_1\right|<x_0<-\left|x_1\right| \\
& \Leftrightarrow\left(x_0, x_1\right) \in D_{\left(-\frac{1}{m}, 0\right), 0}
\end{aligned}
\end{equation*}
\hfill$\blacksquare$\

\begin{remark}\label{rem221} (see \cite{Sto}) If the real coincidence region of an edge-of-the-wedge problem in $\mathbb{C}^2$ consists of two double cones $D_{(-a,0),0}$ and $D_{c,d}$, then one can reach the situation $\left[\left(\frac{1}{a},0\right)+V^{+}\right] \cup \phi\left(D_{c,d}\right) \cup T^{+} \cup T^{-}$ by transforming reciprocal radii. If $\overline{D_{c,d}} \cap \left\{x^2=0\right\} = \emptyset$, then $\phi\left(D_{c,d}\right)$ is again a double cone. With Remark \ref{rem218} (case $\mu=0$), one obtains holomorphic continuation into the set $\widehat{\phi}\left(D_{c,d}\right)$. Undoing the transformation, the lines from $\left(\frac{1}{a},0\right)$ to the endpoints of the double cone $\phi\left(D_{c,d}\right)$ turn into hyperbolas (if they do not have a light-like slope). Thus, in general, one obtains holomorphic extendability into a set $\widehat{D_{c,d}}$ which is bounded by two hyperbolic segments and two light-like line segments. However, if $D_{c,d} \subset \left(D_{(-a,0),0}\right)'$, then by Remark \ref{rem212}, it is certainly true that $\widehat{D_{c,d}} = D_{c,d}$.

No extension is possible if $\overline{D_{c,d}} \subset \left(-\frac{1}{a},0\right) + V^{-}$ or $\overline{D_{c,d}} \subset V^{+}$, since then $\overline{\Phi\left(D_{c,d}\right)} \subset \left(\frac{1}{a},0\right) + V^{-}$.

A detailed discussion of all possible cases will not be carried out here.
\end{remark}

Now, an analogue to Corollary \ref{cor215} will be described for domains $G \subset \mathbb{R}^2$ with the property $G+W=G$; here, $W$ denotes the wedge domain $W:=\left\{x \in \mathbb{R}^2 \mid x^2<0, x_1>0\right\}$.

\begin{theorem}\label{th222}
Let $G \subset \mathbb{R}^2$ be a domain satisfying $G+W=G$, where $W$ is the wedge domain. Let $\widetilde{z} \in \mathbb{R}^2$ be a point lying on a lightlike slope that intersects $\overline{G}$, and let $m \in \mathbb{R}$ satisfy $m \neq 0$ if $\widetilde{z} \in \overline{G}$, or $m<0$ if $\widetilde{z} \notin \overline{G}$. Let $z=\left(z_0, z_1\right) \in \mathbb{C}^2 \backslash \mathbb{R}^2$ be a point satisfying:
\begin{equation*}
(z-\widetilde{z})^2=m
\end{equation*}
Then, $z \in H\left(\widetilde{G} \cup T^{+} \cup T^{-}\right)$. 

Suppose that $a+t b$ with $a, b \in \mathbb{R}^2$ and $t \in \mathbb{R}$, and $b^2=0$ are points on a lightlike line that intersects $\overline{G}$, and let $\tau \neq 0$. Then, $a+(t+i \tau) b \in H\left(\widetilde{W} \cup T^{+} \cup T^{-}\right)$.
\end{theorem} 
\textbf{Proof} We apply the reciprocal radius transformation twice to transform $(\mu, 0) + V^{+}$ to $(0, \mu) + W$, and examine where the points $g(t+i\tau), \tau \neq 0$, of a line $g(t)$ that intersects $(\mu, 0) + V^{+}$ are transformed. According to Corollary \ref{cor215}, such points lie in the holomorphic envelope.\

1st transformation: $\phi$, defined by $w=\phi(x)=-\frac{x}{x^2} ; x=-\frac{w}{w^2}$. It holds that $x \in (k, 0) + V^{+} \Leftrightarrow w \in D_{\left(-\frac{1}{\mu}, 0\right), 0}$ (see Lemma \ref{lem220}).\

2nd transformation: $\psi$, defined by $z=\psi(w)=-\frac{t w-\tilde{w}}{(w-\tilde{w})^2} ; w=-\frac{z}{z^2}+\tilde{w}$ with $\tilde{w}:=\left(-\frac{1}{2 \mu},-\frac{1}{2 \mu}\right)$. Then, $w \in D_{\left(-\frac{1}{\mu}, 0\right), 0} \Leftrightarrow z \in (0, \mu) + W$. The composition $\psi \circ \phi$ yields:
\begin{equation*}
z=-\frac{w-\tilde{w}}{(w-\tilde{w})^2}=-\frac{-\frac{x}{x^2}-\tilde{w}}{\left(-\frac{\pi}{x^2}-\tilde{w}\right)^2}=\frac{\tilde{w} x^2+x}{1+2 x \tilde{w}} .
\end{equation*}
The inverse $(\psi \circ \phi)^{-1}$ is given by:
\begin{equation*}
x=\frac{z-\widetilde{w} z^2}{1-2 z \tilde{w}}.
\end{equation*}
Calculating yields for $|m| \neq 1$ :
\begin{equation*}
x_0=m x_1+c \Leftrightarrow\left(z_0-\frac{c-\mu}{1-m}\right)^2-\left(z_1-\frac{c-m \mu}{1-m}\right)^2=\mu^2 \frac{1+m}{1-m} .
\end{equation*}
Letting $\tilde{z_0}:=\frac{c-\mu}{1-m}, \tilde{z_1}:=\frac{c-m \mu}{1-m}, \lambda:=\mu^2 \frac{1+m}{1-m}$, we obtain:
\begin{equation*}
\begin{gathered}
\tilde{z}_0=\tilde{z_1}-\mu, \
|m|>1 \Leftrightarrow \lambda<0, \
|m|<1 \Leftrightarrow \lambda>0 .
\end{gathered}
\end{equation*}

This means:

1) Lines with timelike slope, which intersect all $(\mu, 0)+V^{+}$, transform into hyperbolas with $\lambda<0$, i.e. such hyperbolas must be taken into account when forming the holomorphic envelope.\par
\hfill \break
2) Lines with spacelike slope intersect $(\mu, 0)+V^{+}$ only if $c>\mu$, i.e. $\tilde{z_0}>0$, thus $\tilde{z} \in(0, \mu)+\overline{W}$, i.e. only $\tilde{z} \in \overline{G}$ need to be considered here.\par
\hfill \break
3) Lines with lightlike slope are transformed back into lines with lightlike slope under $\psi \circ \phi$. Those lines that intersect $(\mu, 0)+V^{+}$ are transformed into those that intersect $(0, \mu)+W$.\par
\hfill \break
According to Corollary \ref{cor215}, all non-real points of lines that intersect $(\mu, 0)+V^{+}$ belong to $H\left(\left((\mu, 0)+V^{+}\right)^{-} \cup T^{+} \cup T^{-}\right)$. However, it holds that $x \in \mathbb{R}^2 \Longleftrightarrow z=\psi \circ \phi(x) \in \mathbb{R}^2$. This means that all points $z \notin \mathbb{R}^2$ that satisfy one of the described hyperbola or line equations lie in $H\left(((0, \mu)+W)^{-} \cup T^{+} \cup T^{-}\right)$.

If the transformations are carried out so that $(-\mu, 0)+V^{-}$ turns into $(0, \mu)+W$, then points $\tilde{z}$ with $\tilde{z}_0=-\tilde{z_1}+\mu$ as hyperbola centers are obtained analogously. These also provide hyperbolas whose non-real points lie in the holomorphic envelope.
This proves the statement. $\hfill\blacksquare$

Another consequence of Corollary \ref{cor215} is the following statement:

\begin{corollary}\label{cor223}
 Let $G \subset \mathbb{R}^2$ contain a double cone $D_{a, b}$. Let $z \in \mathbb{C}^2 \backslash \mathbb{R}^2$ be a point satisfying:
\begin{equation*}
(z-\tilde{x})^2=(b-\tilde{x})^2,
\end{equation*}
where $\bar{x} \in D_{a, b}$ with $(\bar{x}-a)^2>(\bar{x}-b)^2$ (i.e., if $D_{a, b}$ is divided into two halves by the line connecting the remaining two vertices, then $\tilde{x}$ lies in the upper half). Then:
\begin{equation*}
z \in H\left(\tilde{G} \cup T^{+} \cup T^{-}\right),
\end{equation*}
\end{corollary}
\textbf{Proof} To simplify the calculation, we restrict ourselves here to the case from Lemma \ref{lem220}, point 5: $a=\left(-\frac{1}{m}, 0\right), b=0$. Using the transformation of reciprocal radii $\Phi(z)=-\frac{z}{z^2}$, $D_{a, b}$ is biholomorphically mapped to $(m, 0)+V^{+}$. According to Corollary \ref{cor215}, the non-real points of lines intersecting $(m, 0)+V^{+}$ belong to the envelope of holomorphy of $\left((m, 0)+V^{+}\right) \cup T^{+} \cup T^{-}$. It can be verified that with $x=\phi(z)$,
\begin{equation*}
(z-\tilde{x})^2=\widetilde{x}^2 \Longleftrightarrow x_0=\frac{\widetilde{x_1}}{\widetilde{x_0}} x_1-\frac{1}{2 \widetilde{x_0}} .
\end{equation*}
This line intersects the domain $(m, 0)+V^{+}$ when $\left|\frac{\tilde{x}_0}{x_0}\right|<1$ if $-\frac{1}{2 \widetilde{x}_0}>m$, that is, if $\widetilde{x_0}>-\frac{1}{2 m}$. Since this is the case when $\tilde{x}$ lies in the upper half of $D_{a, b}$, the statement is proven. $\hfill\blacksquare$\
\begin{remark}
For symmetry reasons, Corollary \ref{cor223} also holds with the roles of the vertices $a, b$ of the double cone exchanged. In this case, $\tilde{x}$ must be in the lower half of $D_{a, b}$ accordingly.
\end{remark}
With the previous considerations, the preparations are completed to prove the following theorem.\
\begin{theorem}\label{th225}
For all $s \in V^{+}$, it holds that:
\begin{equation*}
H\left(\widetilde{V_m^{+}} \cup\left(\widetilde{D_{0, s}}\right)^{\prime} \cup T^{+} \cup T^{-}\right)=\mathbb{C}^n \backslash\left\{z \mid z^2=\rho \text { for some } \rho \in \mathbb{R} \text { with } 0 \leq \rho \leq m^2\right\} .
\end{equation*}
\end{theorem}
\textbf{Proof} Again, it is sufficient to consider the case $n=2$.

Due to Theorem \ref{th217}, any holomorphic function in $\widetilde{V_m^{+}} \cup\left(\widetilde{D_{0, s}}\right)^{\prime} \cup T^{+} \cup T^{-}$ is also holomorphic in $\bar{R}$.

The statement is proven according to the theorem of Pflug if one can show that any polynomial bounded and holomorphic function in $\widetilde{V_m^{+}} \cup \tilde{R} \cup T^{+} \cup T^{-}$ can be extended holomorphically to $H:=\mathbb{C}^n \backslash\left\{z \mid z^2=\rho\right.$ for some $\rho \in \mathbb{R}$ with $\left.0 \leq \rho \leq m^2\right\}$ (and that there exists a holomorphic function that cannot be extended holomorphically any further).

So let $f$ be holomorphic in $\widetilde{V_m^{+}} \cup \tilde{R} \cup T^{+} \cup T^{-}$, with polynomial bound of order $N$:
\begin{equation*}
    |f(z)| \leq C (\Delta(z))^{-N}
\end{equation*}
with $\Delta(z)=\min \left\{\operatorname{dist}(z) ; \frac{1}{\sqrt{1+|z|^2}}\right\}.$\par
\hfill \break
In $\widetilde{V_m^{+}} \cup \widetilde{R} \cup T^{+} \cup T^{-}$, we have $z^2\neq 0$ everywhere, because for $z=x+i y$, we have:
\begin{equation*}
z^2=0 \Longleftrightarrow x y=0 \text { and } x^2=y^2
\end{equation*}
But this only happens if $x^2=y^2=0$. Such points, however, do not occur in $\widetilde{V_m^{+}} \cup \tilde{R} \cup T^{+} \cup T^{-}$ (see Remark \ref{rem28}, 2).

\begin{figure}[h]
\centering
\includegraphics{"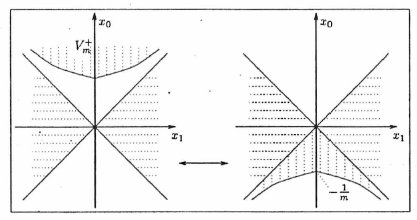"}
\caption{Transformation of reciprocal radii applied to $V_{m}^{+} \cup\left(D_{0, s}\right)^{\prime}$}
\label{Figure 3}
\end{figure}

With the transformation of reciprocal radii, $\widetilde{V_{m}^{+}} \cup \tilde{R} \cup T^{+} \cup T^{-}$ is therefore biholomorphically mapped onto $U(W) \cup U(R) \cup T^{+} \cup T^{-}$, where $W:=\left\{x \in \mathbb{R}^{2} \mid x \in V^{-}, 0<x^{2}<\frac{1}{m^{2}}\right\}$ and $U(W), U(R)$ are certain complex neighborhoods of $W$ and $R$, respectively. Let $\hat{f}:=f \circ \phi$. Then, $\hat{f}$ is holomorphic in $U(W) \cup U(R) \cup T^{+} \cup T^{-}$.

Now, let $w_{0}$ be a point in $\partial V^{-} \backslash{0}$. Choose a neighborhood $V\left(w_{0}\right)$ of $w_{0}$ with the property:

\begin{equation*}
w \in U\left(w_{0}\right) \cap\left(U(W) \cup U(R) \cup T^{+} \cup T^{-}\right) \Rightarrow \text { for } z:=\phi(w) \text { we have } \Delta(z)=\frac{1}{\sqrt{1+|z|^{2}}}.
\end{equation*}

For such $z$, $\frac{1}{\sqrt{1+|\left. z\right|^{2}}}$ should be smaller than the boundary distance. This is possible since, if $U\left(w_{0}\right)$ is chosen small enough, $z$ is far away from the real boundary of $V_{m}^{+}$ or $R$, and because of property 3 in Remark \ref{rem28}, it is also far away from the boundary (in the complex sense) of $\widetilde{V_{m}^{+}} \cup \widetilde{R}$.

Claim: $w \longmapsto \hat{f}(w)\left(w^{2}\right)^{N+1}$ is continuously extendable to $w_{0}$ from $U\left(w_{0}\right) \cap\left(W \cup R \cup T^{+} \cup T^{-}\right)$ with the value 0.

For $w \in E\left(w_{0}\right) \cap\left(W \cup R \cup T^{+} \cup T^{-}\right)$ and $z=\phi(w)$, we have:
\begin{equation*}
\begin{aligned}
\left|\hat{f}(w)\left(w^{2}\right)^{N+1}\right| & =\left|f(z)\left(z^{2}\right)^{-N-1}\right| \leq C(\Delta(z))^{-N}\left|\left(z^{2}\right)^{-N-1}\right|=C\left(\sqrt{1+\|z\|^{2}}\right)^{N}\left|\left(z^{2}\right)^{-N-1}\right| \\
& =C\left(\sqrt{1+\frac{\|w\|^{2}}{\left|w^{2}\right|^{2}}}\right)^{N}\left|\left(w^{2}\right)^{N+1}\right|=C\left(\sqrt{\left(w^{2}\right)^{2}+\|w\|^{2}}\right)^{N}\left|w^{2}\right|.
\end{aligned}.
\end{equation*}

The given equation shows that for any sequence in $U(w_0)\cap(W\cup R \cup T^+ \cup T^-)$ that approaches $w_0$, the expression tends towards zero, since $|w|$ is bounded near $w_0$. This proves the intermediate claim.

Using the Edge-of-the-Wedge Theorem \ref{EOTW}, it follows that $f(w)(w^2)^{N+1}$ is holomorphic in\\  $\left(W \cup R \cup \partial V^{-} \backslash{0}\right)^{\sim} \cup T^{+} \cup T^{-}$. Moreover, by the Jost-Lehmann-Dyson formula \ref{JLD}, $\hat{f}(w)(w^2)^{N+1}$ is also holomorphic in the holomorphic domain $G:=\mathbb{C}^{2} \backslash\left\{w \mid w^{2} \geq \frac{1}{m^{2}}\right\}$.

$f(z)\left(z^{2}\right)^{-N-1}$ is holomorphic in $H$, and hence $f$ is also holomorphic.
$G$ contains $U(W) \cup U(R) \cup T^{+} \cup T^{-}$ (since $H \supset \widetilde{V_{m}^{+}} \cup \tilde{R} \cup T^{+} \cup T^{-}$) and is star-shaped with respect to 0 and therefore simply connected. Thus, the continuation of $\hat{f}(w)\left(w^{2}\right)^{N+1}$ to $G$ is unique, and hence the continuation of $f$ to $H$ is also unique.

Since $G$ is a domain of holomorphy, one can find a function $\tilde{g}$ that is not holomorphically extendable beyond the boundary of $G$. The transformed function $g=\tilde{g} \circ \phi$ cannot be extended beyond the boundary of $H$ (for points $z$ with $z^{2} \neq 0$, this follows directly, and if $g$ were holomorphic at a point $z$ with $z^{2}=0$, it would also be holomorphic in a neighborhood $U(z)$. But since $U(z) \cap\left\{z \mid 0 \leq z^{2} \leq m^{2}\right\} \neq \emptyset$ for all such $z$, this case cannot occur due to the choice of $\tilde{g}$).

Thus, $H$ is a domain of holomorphy and therefore the desired holomorphic envelope. $\hfill\blacksquare$

\subsection{Envelope of holomorphy and hyperboloids}
\begin{equation*}
G_1:=\left\{x \in \mathbb{R}^4 \mid \sqrt{x_1^2+x_2^2+x_3^2+m_1^2}<x_0<\sqrt{x_1^2+x_2^2+x_3^2+m_2^2}\right\} \\
\end{equation*}

The regions $G_1$ and $G_2$ satisfy the conditions of the Jost-Lehmann-Dyson formula \ref{JLD}. Thus, using this formula, one can determine the holomorphic envelopes of $\tilde{G}_1 \cup T^{+} \cup T^{-}$ and $\tilde{G}_2 \cup T^{+} \cup T^{-}$.

Due to the symmetry of $G_1$ and $G_2$ with respect to space rotations, it suffices to calculate the holomorphic envelope in (1+1) dimensions. For spacelike $y$, we denote by $\hat{y}$ as before the vector uniquely determined by $\hat{y}^2=1$, $\hat{y} \in V^{+}$, and $\hat{y} y=0$:
\begin{equation*}
\hat{y}=\frac{\operatorname{sgn} y_1}{\sqrt{-y^2}}\left(y_1, y_0\right)
\end{equation*}

\begin{theorem}\label{th41}
$H\left(\hat{G}_1 \cup T^{+} \cup T^{-}\right)=$\\
$T^{+} \cup T^{-} \cup\left\{z \mid y=0, z \in G_1\right\} \cup\left\{z \mid y^2=0, y \neq 0, x_0>x_1\; \operatorname{sgn} y_0\; \operatorname{sgn}y_1\right\} \cup$\\
$\cup\left\{z=x+i y \mid 0<y^2<\frac{m_2-m_1}{2}\right.$ and $\left.F^{-}\left(x_1, y\right)<x_0<F^{+}\left(x_1, y\right)\right\}$ with\

\begin{equation*}
\begin{aligned}
F^{-}\left(x_1, y\right) & :=-\hat{y}_0 \sqrt{\left(\frac{m_2-m_1}{2}\right)^2+y^2}+\sqrt{\left(\frac{m_2+m_1}{2}\right)^2+\left(x_1+\hat{y}_1 \sqrt{\left(\frac{m_2-m_1}{2}\right)^2+y^2}\right)^2} \\
F^{+}\left(x_1, y\right) & :=\hat{y}_0 \sqrt{\left(\frac{m_2-m_1}{2}\right)^2+y^2}+\sqrt{\left(\frac{m_2+m_1}{2}\right)^2+\left(x_1-\hat{y}_1 \sqrt{\left(\frac{m_2-m_1}{2}\right)^2+y^2}\right)^2}.
\end{aligned}
\end{equation*}
\end{theorem}
In the case where $y^2<0$, this theorem states that for a fixed $y$, $x$ lies between the upper branches of the two hyperbolas $\left(x \pm \hat{y} \sqrt{\left(\frac{m_2-m_1}{2}\right)^2+y^2}\right)^2=\left(\frac{m_2+m_1}{2}\right)^2$.

\textbf{Proof} The Jost-Lehmann-Dyson formula \ref{JLD} is used. It is
\begin{equation*}
N\left(G_1\right)=\left\{\left(x^{\prime}, \lambda\right) \mid x^{\prime} \in \overline{V^{+}}, \lambda \geq \max \left\{m_2-\sqrt{x^{\prime 2}}, \sqrt{x^{\prime 2}}-m_1\right\}\right.
\end{equation*}
With this we now calculate the boundary of $H\left(\tilde{G_1} \cup T^{+} \cup T^{-}\right)$.\\\
$\left(z-x^{\prime}\right)^2=\lambda^2$ means:\\
(1) $\left(x-x^{\prime}\right)^2-y^2=\lambda^2$\\
(2) $\left(x-x^{\prime}\right) y=0$.\\

From (2) it follows that $x-x^{\prime}=\mu \hat{y}, \mu \in \mathbb{R}$.
Using (1) we get $\mu^2-y^2=\lambda^2$, hence $\mu= \pm \sqrt{\lambda^2+y^2}$.
If $x^{\prime}=a \hat{y}+b y\left(\Rightarrow x^{\prime 2}=a^2+b^2 y^2\right)$, then we obtain with $\alpha:=\sqrt{x^{\prime 2}}$:
\begin{equation*}
x=\left( \pm \sqrt{\lambda^2+y^2}+\sqrt{\alpha^2-b^2 y^2}\right) \hat{y}+b y .
\end{equation*}
If $y$ is given, then the \glqq +\grqq{} sign describes the points above the holomorphic envelope (i.e. with larger $x_0$ values); the \glqq -\grqq{} sign corresponds to points below the holomorphic envelope. Eliminating the parameter $b$ from the last equation yields (in the case of \glqq +\grqq{}):
\begin{equation*}
x_0=\hat{y}_0 \sqrt{\lambda^2+y^2}+\sqrt{\alpha^2+\left(x_1-\hat{y}_1 \sqrt{\lambda^2+y^2}\right)^2} .
\end{equation*}
where:

\begin{equation*}
\alpha \geq 0,\quad \lambda \geq \max \left\{m_2-\alpha, \alpha-m_1\right\}.
\end{equation*}

$x_0$ is strictly monotonically increasing in $\alpha$, as differentiation shows: $\frac{\partial x_0}{\partial \lambda}>0$, because the derivative of the first term of $x_0$ is positive and has a magnitude greater than that of the second term. This means that by decreasing $\alpha$ and $\lambda$ while holding $x_1$ and $y$ constant, we obtain smaller $x_0$-values. Thus, the smallest $x_0$-value is obtained when $\lambda=m_2-\alpha$ and $0 \leq \alpha \leq \frac{m_1+m_2}{2}$. Substituting this value, we get:
\begin{equation*}
x_0=\hat{y}_0 \sqrt{\lambda^2+y^2}+\sqrt{\left(m_2-\lambda\right)^2+\left(x_1-\hat{y}_1 \sqrt{\lambda^2+y^2}\right)^2}.
\end{equation*}
Which value of $\lambda$, $\frac{m_2-m_1}{2} \leq \lambda \leq m_2$, now yields the smallest value of $x_0$?
\begin{equation*}
\frac{\partial x_0}{\partial \lambda}=\frac{\lambda}{\sqrt{\lambda^2+y^2}} \hat{y}_0+\frac{\lambda-m_2+\left(x_1-\hat{y}_1 \sqrt{\lambda^2+y^2}\right) \hat{y}_1 \frac{-\lambda}{\sqrt{\lambda^2+y^2}}}{\sqrt{\left(m_2-\lambda\right)^2+\left(x_1-\hat{y}1 \sqrt{\lambda^2+y^2}\right)^2}} .
\end{equation*}
For which values of $x{1}, y, \lambda$ is this expression positive?

Upon calculation, we find that $\frac{\partial x_{0}}{\partial \lambda}>0 \Longleftrightarrow$

$x_{1}^{2}-2 \frac{m_{2} \hat{y_{1}}}{\lambda} \sqrt{\lambda^{2}+y^{2}}\, x_{1}+\left(m_{2}-\lambda\right)^{2} \hat{y}_{0}^{2}+\left(\lambda^{2}+y^{2}\right)\left(\frac{2 \hat{y}_{1}^{2} m_{2}}{\lambda}-\left(\hat{y}_{0}-\frac{m_{2}}{\lambda}\right)^{2}\right)>0$.

This is a quadratic expression in $x_{1}$. It describes an upward-opening parabola with fixed remaining parameters, and has, as can be calculated, no roots.

Therefore, $\frac{\partial x_{0}}{\partial \lambda}>0$ always holds. So the smallest value of $\lambda=\frac{m_{2}-m_{1}}{2}$ yields the boundary.

\begin{equation*}
x_{0}=\hat{y}_0 \sqrt{\left(\frac{m_{2}-m_{1}}{2}\right)^{2}+y^{2}}+\sqrt{\left(\frac{m_{2}+m_{1}}{2}\right)^{2}+\left(x_{1}-\hat{y}_{1} \sqrt{\left(\frac{m_{2}-m_{1}}{2}\right)^{2}+y^{2}}\right)^{2}}.
\end{equation*}

The same procedure is applied for the lower boundary in the range $y^{2}<0$. The claim is also obtained there.\\

For the case $y^{2}=0, y \neq 0$, one has to take the closure of those points $z$ with $y^{2}<0$ that do not lie in $H\left(\tilde{G}_{1} \cup T^{+} \cup T^{-}\right)$.

For $z \in H\left(\tilde{G}{1} \cup T^{+} \cup T^{-}\right)$, $x$ lies precisely between the upper branches of the hyperbolas describing the boundary, given by
\begin{equation*}
\left(x \pm \hat{y} \sqrt{\left(\frac{m_{2}-m_{1}}{2}\right)^{2}+y^{2}}\right)^{2}=\left(\frac{m_{2}+m_{1}}{2}\right)^{2}.
\end{equation*}

For $y_{1} \longrightarrow y_{0} \neq 0$, this yields $x_{0} \leq x_{1}$ sgn $y_{0}$ sgn $y_{1}$.

This proves the claim.

For $y=0$, one obtains the desired result directly from the definition of $N(G)$. \hfill$\blacksquare$\par
\hfill\break

\section{Application of holomorphic continuation to a mass gap situation}\label{Mass gap}
With the preparations from the previous sections, we can now prove the following statement:

\begin{theorem}\label{th315}
Let $\{\mathcal{A}(O), \mathcal{A}, M, \alpha\}$ be a theory of local observables with the covariant factorization representation $\pi$. For the minimal $U$ corresponding to $\pi$, spec $U$ cannot be restricted in the following way:

$$\operatorname{spec}\, U \subset\left\{p \mid m^{2} \leq p^{2} \leq m_{1}^{2}, p_{0}>0\right\}, \quad m<m_{1}<\infty$$

where $m$ is chosen maximally, i.e., $\left\{p \mid p^{2}=m^{2}, p_{0}>0\right\} \subset \operatorname{spec} U$.
\end{theorem}

\textbf{Proof} We assume that the spectrum of $U$ is localized as follows:
\begin{equation*}
\text{spec } U \subset\left\{p \mid m^2 \leq p^2 \leq m_1^2, p_0>0\right\} \subset V^{+}
\end{equation*}
and the spectrum starts exactly at $m$, and we will use these assumptions to arrive at a contradiction.

For $t \in V^{+}$, we choose $x \in \mathcal{A}\left(D_{-t, t}\right)$ and define the functions:
\begin{equation*}
\begin{aligned}
& F_{x, \psi}^{+}(a)=\left(\Psi, \pi(x^*) U(a) \pi(x) \Psi\right) \text { and } \\
& F_{x, \psi}^{-}(a)=\left(\Psi, \pi(\alpha_a x) \pi(x^*) U(a) \Psi\right)=\left(\Psi, E(a) \pi(x) U(-a) \pi(x^) U(a) \Psi\right) .
\end{aligned}
\end{equation*}
Due to Axiom 2 (locality), we have:
\begin{equation*}
F_{x, \Psi}(a):=F_{x, \psi}^{+}(a)-F_{x, \psi}^{-}(a)=0, \text { if } a \in\left(D_{-2 t, 2 t}\right)^{\prime}
\end{equation*}
since $\pi\left(\alpha_a x\right)$ and $\pi\left(x^*\right)$ commute in this case.

Therefore, since supp $F_{x, \psi} \subset\left(-2 t+V^{+}\right) \cup\left(2 t+V^{-}\right)$, we can split $F_{x, \psi}$ into $F_{x, \psi}=G^{+}-G^{-}$, such that
\begin{equation*}
\begin{aligned}
& \operatorname{supp} G^{+} \subset-2 t+V^{+} \
& \operatorname{supp} G^{-} \subset 2 t+V^{-} .
\end{aligned}
\end{equation*}
Outside of $D_{-2 t, 2 t}$, this splitting is unique.\\

Due to Theorem \ref{th26}, the following holds for the Fourier transforms:

$\widetilde{G^{+}}$ is the boundary value in the distributive sense of a holomorphic function in $T^{+}$, which is also denoted by $\widetilde{G^{+}}$. Similarly, $\widetilde{G^{-}}$ is the boundary value of a holomorphic function in $T^{-}$.
Furthermore, for real points, the following holds:
\begin{equation*}
\widetilde{G^{+}}(p)=\widetilde{G^{-}}(p) \quad \forall p \in \Gamma:=M \backslash \operatorname{supp} \widetilde{F_{x, \psi}} .
\end{equation*}
The Edge-of-the-Wedge Theorem \ref{EOTW} is tailored to this situation: There exists a holomorphic function $G^*$ on $T^{+} \cup T^{-} \cup \tilde{\Gamma}$ such that:
\begin{equation*}
\begin{gathered}
G^*\left|T^{+}=\widetilde{G^{+}},\; G^{*}\right| T^{-}=\widetilde{G^{-}} \text { and } \
G^+(p)=\widetilde{G^{+}}(p)=\widetilde{G^{-}}(p) \quad \text {for } p \in \Gamma .
\end{gathered}
\end{equation*}
Due to Lemma \ref{lem316}, further statements can be made about the shape of the region $\Gamma$: $\operatorname{supp} \widetilde{F_{x, \Psi}} \subset \operatorname{supp} \overline{F_{x, \Psi}^{+}} \cup \operatorname{supp} \widetilde{F_{x, \Psi}^{-}} \subset \operatorname{spec} U \cup(2 S-\operatorname{spec} U)$, where $S$ denotes the support of $\Psi$.

\begin{figure}[h]
 \centering
 \includegraphics{"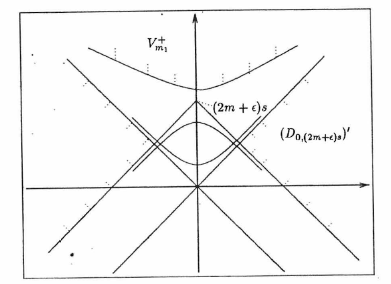"}
 \caption{Portion of the coindidence domain contained in  $\Gamma$}
 \label{Abbildung 4}
\end{figure}

If $S$ is a small Borel set containing $m s$ (where $s$ is a vector in $V^{+}$ with $s^2=1$), then we obtain:

$\Gamma \supset\left\{p \mid p^2>m_{\mathbf{1}}^2, p_0>0\right\} \cup\left(D_{0,(2 m+\epsilon) s}\right)^{\prime} \quad$ for a small $\epsilon$ determined by $\mathrm{S}$.

Using Theorem \ref{th225}, we see that $G^*$ is also holomorphic in $N:=M \backslash\left\{p \mid 0 \leq p^2 \leq m_1^2\right\}$. Since $\widetilde{G^{+}}$ and $\widetilde{G^{-}}$ are boundary values of $G^*$ at real points, we have
\begin{equation*}
\widetilde{G^{+}}(p)=\widetilde{G^{-}}(p) \quad \forall p \in N
\end{equation*}
Hence, we also have: $\widetilde{F_{x, \Psi}} \mid N=0$.
Since supp $\widetilde{F_{x, \psi}^{+}} \cap N=\emptyset$, it follows that $\widetilde{F_{x, \psi}^{-}} \mid N=0$.

Claim: $q \in 2 m s-\left\{p \mid p^2=m^2, p_0>0\right\} \Rightarrow \exists \Psi_0$ with supp $\Psi_0 \ni m s$ and $x_0 \in \mathcal{A}\left(D_{-t, t}\right)$ for a large $t$, such that $q \in \operatorname{supp} \widetilde{F_{x_0, \Psi_0}^{-}}$.\\
Proof of this claim: For any neighborhood $V$ of $m s$, $E(V) \neq 0$; therefore, by Lemma \ref{lem317}, $\overline{K(V)}=\mathcal{H}$. Hence,
\begin{equation*}
K^{\prime}(V):=\{\Phi \mid \Phi=\int_V \pi\left(x^*\right) d E(p) \Psi, \,\pi\left(x^*\right) \Psi \in K(V)\}
\end{equation*}
is dense in $\mathcal{H}$. It follows that for any relatively compact $W$ with $W \cap$ spec $U \neq \emptyset$, there exists $\Phi \in K^{\prime}(V)$ such that
\begin{equation*}
\int_W(\Phi, d E(p) \Phi) \neq 0;
\end{equation*}
because if it were zero for all $\Phi \in K^{\prime}(V)$, then it would also be zero for all $\Phi \in \mathcal{H}$, but that would imply $W \cap \text{spec } U = \emptyset$. Now, if $q \in 2ms - \{p \mid p^2 = m^2, p_0 > 0\}$, then for every relatively compact neighborhood $U(q)$, there exists an $S \ni ms$ and a relatively compact $W$ with $W \cap \text{ spec } U \neq \emptyset$ such that $2S - W = U(q)$. Furthermore, there exist $\Phi \in K^{\prime}(S)$ and $\rho \in \mathcal{S}(M)$ with $\rho \mid U(q) \equiv 1$, such that
$$
\int_W \rho(p)(\Phi, d E(p) \Phi) \neq 0 .
$$
Write $\Phi$ as $\Phi=\int_S \pi\left(x_0^*\right) d E(s) \Psi_0$, then it is shown that $q \in \operatorname{supp} \widetilde{F_{x_0, \Psi_0}^{-}}$, since $U(q)$ can be chosen arbitrarily small. This proves the intermediate claim.

Since $2 m s-\left\{p \mid p^2=m^2, p_0>0\right\} \cap N \neq \emptyset$, a contradiction to the location of the support of $\widetilde{F_{x_0, \psi_0}^{-}}$ is obtained. This proves the theorem. $\hfill\blacksquare$

\section{Outlook}\label{outlook}
With the formulation of the millennium problems in \cite{JW}, the topic of the mass gap became more visible again. Although in \cite{Bor6} it is shown that a theory without mass gap is consistent with the axiomatic approach, the techniques elaborated in this article might still prove useful in that context. Relevant literature that combines methods of holomorphic continuation with the millennium problem and Yang-Mills-Theory seems to be rare though.\\

Another interesting application for the techniques described in this article is the following: 
Special representations of the observable algebra are the factor representations, also known as superselection sectors. They represent the set of states with the same charge quantum numbers.

For example, consider the state that corresponds to the presence of a particle with a specific charge. The state corresponding to two of these particles plus a corresponding antiparticle again has the same charge. However, the possible energy-momentum values for the 3-particle system should belong to the same superselection sector again. Denoting the spectrum of translations to superselection sector A as $S_A$, one would thus conjecture that (s. \cite{Bor5}):

\begin{equation*}\label{conjecture1}
3 \; S_A \subset S_A 
\end{equation*}

A mathematical proof of this conjecture would be desirable. 
It is likely that it can be proven with the techniques of this article.

\addcontentsline{toc}{section} {Glossary}
\section*{Glossary}
Table of commonly used symbols \\
\renewcommand{\arraystretch}{1.5}
\begin{tabular}{lp{12cm}r}
 & &   Seite \\ 

$M$ & Minkowski space & \pageref{minkowskispace} \\
$V^{+}$ & Forward light cone & \pageref{forwardcone} \\
$V^{-}$ & Backward light cone & \pageref{backwardcone} \\
$T^{+}$ & $=\mathbb{R}^n+i V^{+}$, forward tube & \pageref{forwardtube} \\
$T^{-}$ & $=\mathbb{R}^n+i V^{-}$, backward tube & \pageref{backwardtube} \\
$V_\mu^{+}$ & $\left\{x \in V^{+} \mid x^2>\mu^2\right\}$ & \pageref{Vmuplus} \\
$R$& Set of spacelike points & \pageref{spacelike} \\
$S^{\prime}$ & ($S \subset \mathbb{R}^n$), set of spacelike points corresponding to each $x \in S$ & \pageref{spacelikeSet} \\
$N(.)$ & Set of parameters for admissible hyperbolas or hyperboloids & \pageref{admissibleHyp} \\
$N_{\infty}(.)$ & Set of parameters for admissible lines or planes & \pageref{admissibleinfinity}\\
$\mathcal{N}^{\prime}$ & $(\mathcal{N} \subset \mathcal{B}(\mathcal{H}))$, commutant of $\mathcal{N}$ & \pageref{Kommutante} \\
$D_{a, b}$ & Double cone spanned by $a, b \in \mathbb{R}^n$ & \pageref{double cone} \\
$H(G)$ & Envelope of holomorphy of the domain $G \subset \mathbb{C}^n$ & \pageref{HolomorphicEnvelope} \\

$\tilde{B}$ & ($B \subset \mathbb{R}^n$ region), neighborhood of $B$ resulting from the Edge-of-the-Wedge-Theorem & \pageref{tildeB} \\

$f$ & (f function), Fourier transform of $f$ & \pageref{Fouriertransformierte}\\
\end{tabular}

\addcontentsline{toc}{section} {Table of figures}
\listoffigures

\addcontentsline{toc}{section} {References}

\bibliographystyle{alpha}
\bibliography{ConstructionofenvelopesofholomorphyandQFT}

\hfill \break
Ulrich Armbrüster, Vienna/Austria; previous affiliation: University of Göttingen, Mathematisches Institut\\
\href{mailto:armbruester@gmx.net}{armbruester@gmx.net}

\end{document}